\documentclass{optica-article}
\journal{opticajournal} % for journals or Optica Open
\articletype{Research Article}
\usepackage{lineno}
\begin{document}

\title{Phase retrieval based on intensity-only spatiotemporal wavefront shaping}
\author{Luis Alberto Razo-L\'opez,\authormark{1,*} Marc Guillon,\authormark{1,2} Fabrice Lemoult,\authormark{1} and S\'ebastien M. Popoff\authormark{1}}
\address{\authormark{1}Institut Langevin, ESPCI Paris, Universit\'e PSL, CNRS, Paris 75005, France\\
\authormark{2}Saint-P\`eres Paris Institute for Neurosciences (SPPIN), CNRS, Paris, France}
\email{\authormark{*}luis-alberto.razo-lopez@espci.fr}

\begin{abstract*}
We introduce a phase retrieval method based on harmonic field representations generated by intensity-only spatiotemporal modulation. 
A periodic amplitude angular modulation encodes phase information into temporal intensity harmonics, 
enabling wavefront reconstruction without interferometry, reference beams, or a calibrated phase modulator. 
We consider a disordered system that can be modeled as a phase plate and measure the intensity pattern 
in the far field for a rotating excitation pattern generated by a digital micromirror device (DMD). 
We extract the harmonic components of the spatiotemporal speckle and exploit 
a gradient-based optimization algorithm to retrieve both the optical phase 
in the plane of the camera corresponding to a plane-wave illumination 
and the effective phase plate that represents the disordered system. 
We demonstrate experimentally the accuracy of the approach in quantitatively estimating 
and compensating for phase distortion ranging from low-spatial-frequency aberrations 
to high-frequency distortions of a ground-glass diffuser. 
The proposed approach establishes a new route toward calibration-free phase retrieval using intensity-only measurements.
\end{abstract*}

\section{Introduction}

Recovering phase information at the wavelength scale from intensity-only measurements is a central problem in optical imaging~\cite{Shechtman_2015, Jaganathan_2015, Zhang_2023}, with applications ranging from microscopy~\cite{Lee_2014, Wang_2019} and wavefront sensing~\cite{Zhao_2024} to imaging through scattering media~\cite{Lei_2024}.
Conventional approaches to phase retrieval rely on interferometric detection~\cite{Lee_2014, Duadi_2011, Venturi_2017}, holographic references~\cite{Barmherzig_2019, Stockmar_2013}, or iterative algorithms such as Gerchberg–Saxton~\cite{Gerchberg_1972} and Fienup methods~\cite{Fienup_1982}.
Although powerful, these techniques typically require precise calibration, stable reference beams, or multiple controlled measurements.
Alternative strategies based on intensity-only measurements have been extensively explored, including transport-of-intensity methods~\cite{Teague_1983} and modern optimization-based approaches~\cite{Candes_2015}.
Numerical and statistical-inference frameworks have further broadened the scope of intensity-only phase retrieval, in particular Bayesian and machine-learning methods, which have been applied to the reference-less estimation of the transmission matrix of strongly scattering media~\cite{Dremeau_2015, Metzler_2017}.
In parallel, techniques such as ptychography have demonstrated high-fidelity phase reconstruction through structured measurement diversity~\cite{Rodenburg_2008}.
However, these approaches often depend on strong priors, known system transfer functions, or multiple measurement planes, limiting their applicability in complex or poorly characterized optical systems.

Imaging through scattering media has also been addressed using wavefront shaping and transmission matrix approaches~\cite{Vellekoop_2007, Popoff_2010}, enabling control of light propagation in complex media.
Recent advances in spatiotemporal wavefront shaping have opened new possibilities for encoding spatial information into temporal signals~\cite{Noetinger_2023, Noetinger_2024, Jia_2025}.
In particular, structured temporal modulation enables indirect control of phase modulation through intensity fluctuations.

In this work, we build on these concepts and introduce a phase retrieval framework based on a rotating angular modulation combined with a thin scattering diffuser.
We show that the temporal Fourier decomposition of the measured intensity signal yields harmonics that correspond to spiral-phase modulated spatial modes, establishing a direct mapping between temporal frequency and spatial phase structure.
Based on the harmonic structure induced by the modulation, we formulate a calibration-free inverse problem that reconstructs both the far-field wavefront and the near-field phase.
The method requires only intensity images and does not rely on prior knowledge of the optical transfer function, making it applicable to partially unknown or aberrated optical systems.
From an experimental perspective, our approach leverages a digital micromirror device (DMD) to implement high-speed intensity modulation.
Compared to phase-only spatial light modulators (SLMs), DMDs offer significantly higher refresh rates (typically up to few tens of kHz), enabling rapid acquisition of the temporal sequences required for harmonic analysis~\cite{Hornbeck_1995, Dudley_2003, Akbulut_2011}. 
 
\section{Spatiotemporal encoding of the spatial phase information}

\subsection{Principle}

\begin{figure}
\centering\includegraphics{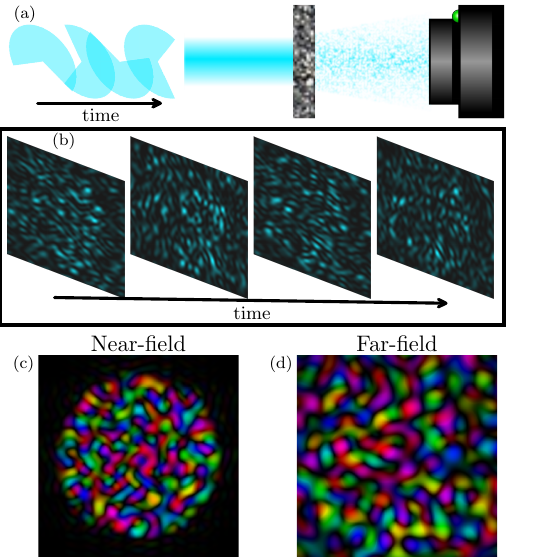}
\caption{\textbf{Principle of spatiotemporal phase encoding.}
(a) Schematic of the experimental concept. A coherent beam is modulated by a time-dependent angular aperture $M(r,\theta,t)$ and propagates through a thin diffuser.
(b) The set of time-varying speckle patterns recorded by a camera.
From the set of intensity measurements of (b) we can retrieve both the phase profile of the thin diffuser (c) as well as the phase of the speckle intensity pattern (d) obtained under uniform static illumination of the diffuser in (c).}
\label{fig:SpatioTemp}
\end{figure}

The proposed method is illustrated in Fig.~\ref{fig:SpatioTemp}.
A coherent beam is intensity-only modulated by a time-dependent angular mask and propagates through a thin static phase diffuser before being recorded by a camera [see Fig.~\ref{fig:SpatioTemp}(a)].
The modulation consists of the full pupil from which an angular sector of width $\alpha$ has been removed; this dark sector rotates at angular velocity $\Omega$~\cite{Noetinger_2023, Noetinger_2024, Jia_2025} (the explicit expression of the mask is given in Section~S1 of the Supplementary Material).
A representative diffuser phase profile and the static speckle it produces in the far field under uniform illumination are shown in Figs.~\ref{fig:SpatioTemp}(c) and (d), respectively.

Because the illumination rotates, the otherwise static speckle becomes a temporally periodic signal: at each camera pixel the recorded intensity oscillates in time [see Fig.~\ref{fig:SpatioTemp}(b)], as different angular sectors of the pupil successively probe the diffuser.
The method exploits this conversion of spatial phase into temporal intensity fluctuations and proceeds in three steps.
First, the temporal intensity recorded at every pixel is Fourier transformed, producing a discrete set of harmonics of the rotation frequency $\Omega$.
Second, for a suitable sector angle these harmonics reduce to a few observables that depend on the interference between spiral-phase–modulated copies of the field, thereby encoding the otherwise inaccessible phase.
Third, a gradient-descent optimization fits these observables with a forward model 
and retrieves both the complex field in the camera (far-field) plane 
and the phase of the diffuser in the near-field plane, 
without any reference beam, phase modulator, or prior calibration of the system.
We show examples of such reconstructed fields in Fig.~\ref{fig:SpatioTemp}(c) and (d).

\subsection{Temporal Fourier decomposition}

We first establish the harmonic content generated by the rotating mask.
Since the modulation is periodic with period $T=2\pi/\Omega$, the complex field reaching the detector can be expanded as a temporal Fourier series,
\begin{equation}
    h(r,\theta,t) = \sum_{n=-\infty}^{\infty} c_n\, a_n(r,\theta)\, \mathrm{e}^{\mathrm{i}n \Omega t}\,,
    \label{eq:comExpand}
\end{equation}
with
\begin{equation}
    c_n = \mathrm{sinc}\!\left(\frac{n \alpha}{2}\right)\mathrm{e}^{-\mathrm{i}n \phi_0}\,,
\end{equation}
where $a_n(r,\theta,t)$ is the normalized spatial field associated with the $n$-th harmonic, 
$c_n$ are the Fourier coefficients of the rotating mask, and $\phi_0$ is its initial angular offset (the full derivation is given in Section~S1 of the Supplementary Material).

\begin{figure}
\centering\includegraphics{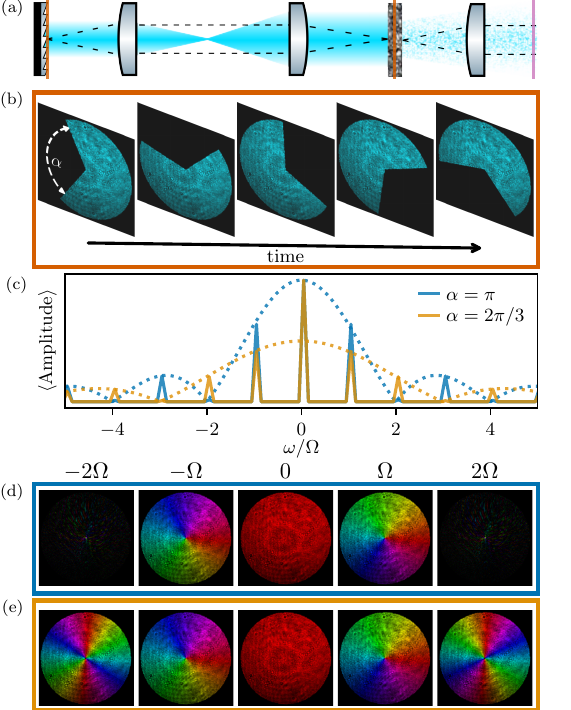}
\caption{\textbf{Experimental  harmonic structure and spiral-phase decomposition.}
(a) Schematic view of the experimental configuration used to generate and measure the spatiotemporal modulation. Color lines indicate: the near-field plane (orange) and the far-field plane (pink).
(b) Temporal intensity signal recorded by the camera.
(c) Fourier spectrum of the measured intensity signal. Solid lines correspond to experimental measurements for $\alpha=\pi$ (blue) and $\alpha=2\pi/3$ (yellow). Dashed lines indicate the corresponding envelopes $\mathrm{sinc}(n\alpha/2)$.
(d) Spatial complex images of the first five harmonics for $\alpha=\pi$.
(e) Same as (d) for $\alpha=2\pi/3$.}
\label{fig:Modulation}
\end{figure}

Two complementary aspects govern this decomposition.
First, $a_n$ corresponds to the field obtained when the pupil is modulated  by a spiral phase $\mathrm{e}^{\mathrm{i}n\theta}$,  each temporal harmonic carrying a well-defined orbital angular momentum,  with topological charge equal to its order $n$.
Second, the harmonic amplitudes are weighted by the envelope $\mathrm{sinc}(n\alpha/2)$, set by the angular width $\alpha$, which selectively suppresses specific harmonic orders. 
We validate this structure experimentally [Fig.~\ref{fig:Modulation}].
A narrowband polarized laser beam is expanded and modulated by a DMD (see Section~S2 of the Supplementary Material for the setup) to generate the rotating mask, which is imaged onto the near-field plane through a $4f$ system [orange lines in Fig.~\ref{fig:Modulation}(a)] and recorded by a camera [Fig.~\ref{fig:Modulation}(b)].

Because the modulated field is imaged directly, the temporal Fourier transform of the signal at each pixel gives access to the linear coefficients $c_n$ and yields the harmonic comb of Fig.~\ref{fig:Modulation}(c).
Its envelope follows $\mathrm{sinc}(n\alpha/2)$: for $\alpha=\pi$ all even harmonics vanish (blue curve), whereas for $\alpha=2\pi/3$ every third harmonic is suppressed (yellow curve).
The spatial images of the first harmonics [Figs.~\ref{fig:Modulation}(d) and (e)] confirm the associated spiral-phase modes: the zeroth order is a uniform field and each order $n\neq0$ a vortex of topological charge $n$, while the orders cancelled by the envelope appear as vanishing fields.

\subsection{Intensity harmonics for $\alpha=\pi$}

In the phase-retrieval configuration the field has propagated through the diffuser, and only the speckle intensity $H(r,\theta,t)$ is accessible,
\begin{equation}
    H(r,\theta,t) = \left|\sum_{n} c_n a_n(r,\theta)\, \mathrm{e}^{\mathrm{i}n \Omega t}\,\right|^2. \label{eq:absExpand}
\end{equation}
This quadratic relation couples harmonic components.
The choice $\alpha=\pi$ makes it tractable: the even orders vanish and the envelope decays quickly, so the field is well approximated by retaining only the terms $\left|n\right|\leq 1$,
\begin{equation}
    h(t) \approx c_0a_0(r,\theta)  + c_1a_1(r,\theta)\mathrm{e}^{\mathrm{i}\Omega t}+c_{-1}a_{-1}(r,\theta)\mathrm{e}^{-\mathrm{i}\Omega t}\,. \label{eq:appoxExpand}
\end{equation}
The temporal Fourier decomposition of the measured intensity then retains only a few non-zero harmonics, of which the relevant ones are
\begin{equation}
    H(r,\theta,t) \approx H_0(r,\theta)  + H_1(r,\theta) \mathrm{e}^{\mathrm{i}\Omega t} + H_{-1}(r,\theta)  \mathrm{e}^{-\mathrm{i}\Omega t}\,, \label{eq:HarmsExpand}
\end{equation}
with
\begin{equation}
    \begin{aligned}
        H_0(r,\theta) &= |a_0(r,\theta)|^2+c_1^2 \left(|a_1(r,\theta)|^2+|a_{-1}(r,\theta)|^2\right), \\
        H_1(r,\theta) &= H_{-1}^*(r,\theta) = c_1a_0^*(r,\theta)a_1(r,\theta) + c_1^*a_0(r,\theta)a_{-1}^*(r,\theta),
    \end{aligned}
    \label{eq:Harms}
\end{equation}
where $^*$ denotes complex conjugation.
Importantly, the first harmonic $H_1$ carries the interference between the fundamental mode $a_0$ and the first-order spiral modes $a_{\pm1}$, thereby encoding the phase information within intensity-only images.
Together with the constant term $|a_0|^2$, 
obtained directly from the intensity measurement under a static disk illumination, 
the three observables $|a_0|^2$, $H_0$ and $H_1$ constitute the dataset 
used by the reconstruction algorithm.

\section{Phase retrieval algorithm}

\begin{figure}
\centering\includegraphics{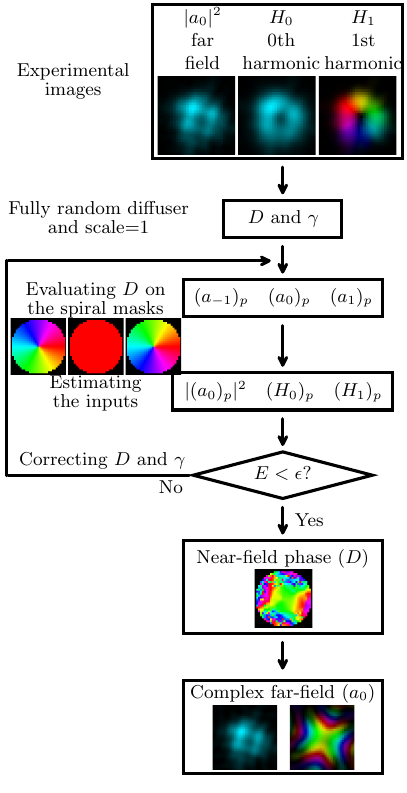}
\caption{\textbf{Optimization-based phase retrieval algorithm based on harmonic decomposition.}
The experimental inputs extracted from temporal measurements, consisting of the static intensity $|a_0|^2$, $H_0$ and $H_1$ (amplitude and phase).
The diffuser phase $D$ is initialized randomly and sequentially modulated by spiral-phase masks of orders $n=0$ and $\pm1$, generating predicted quantities $|(a_0)_p|^2$, $(H_0)_p$ and $(H_1)_p$. 
A global scaling parameter $\gamma$ is applied to account for unknown system sampling (see Section~S3 of the Supplementary Material).
The discrepancy between measured and predicted quantities is minimized using a normalized $L^2$ loss and the a gradient-based Adam optimizer is used to update the diffuser phase to minimize this loss.
The diffuser phase and scaling parameter are updated until convergence $(E<\epsilon)$.
Inset panels illustrate representative examples of the quantities involved in the reconstruction (inputs, spiral masks, and reconstructed fields).
}
\label{fig:Algorithm}
\end{figure}

The objective is to recover the unknown diffuser phase $D$, 
and its resulting far-field in the plane of the camera, from the set of intensity-only measurements.
The reconstruction procedure is summarized in Fig.~\ref{fig:Algorithm}.
% For illustration purposes, representative examples of the quantities involved in the reconstruction are shown within the diagram. 
% These include the input observables $|a_0|^2$, $H_0$, and $H_1$, the spiral-phase masks $\mathrm{e}^{\mathrm{i}n\theta}$, and the reconstructed near- and far-fields $D$ and $a_0$, respectively.
% These panels are provided as visual guides and are not used directly in the optimization process.
The algorithm takes as input three experimentally measured quantities: $|a_0|^2$, $H_0$ and $H_1$.
These quantities are extracted from a single temporal acquisition via temporal Fourier decomposition.
The reconstruction is then initialized with a fully random phase distribution for the diffuser $D$.
At each iteration, the current diffuser estimate feeds the forward model: the far fields corresponding to $D$ modulated by spiral-phase masks of orders $n=0$ and $\pm1$ are computed and combined through Eqs.~\eqref{eq:Harms} to predict the quantities $|(a_0)_p|^2$, $(H_0)_p$ and $(H_1)_p$, where the subscript $p$ denotes the model predictions.
The optimization additionally adjusts a global scaling parameter that compensates for the unknown spatial sampling 
of the system without explicit calibration (see Section~S3 of the Supplementary Material), making the method applicable to systems with partially characterized transfer functions, including partially developed speckle regimes or aberrations.

The discrepancy between measured and predicted quantities is quantified using a normalized $L^2$ loss defined as:
\begin{equation}
    E = \sum_{B\in \{|a_0|^2, H_0, H_1\}}\sqrt{\frac{\langle|B-B_p|^2\rangle}{\langle|B|\rangle}}. \label{eq:Error}
\end{equation}
This formulation ensures balanced contributions from all observables, independently of their absolute intensity scale.
The loss function is minimized using the Adam optimizer implemented in PyTorch~\cite{Kingma_2017}, 
and the iterative loop continues until the reconstruction error satisfies a predefined convergence criterion $E<\epsilon$, 
or until a maximum number of iterations is reached.
To avoid local minima in the high-dimensional parameter space the optimization follows a multiscale strategy. 
The idea is to first recover a coarse representation of the diffuser phase and progressively refine it to full resolution.
This allows the algorithm to focus on smooth solutions, avoiding non-physical solutions with discontinuous phase jumps and spatial frequency components forbidden by the system's numerical aperture (See Section S4 of the Supplementaty Material).

Phase retrieval from intensity-only measurements is always subject to inherent ambiguities. 
In the present formulation, the recovered phase admits two conjugate solutions corresponding to the diffuser phase $D(r)$ and its inverted conjugate $D(-r)^*$, which produce identical intensity measurements.
In our approach, this ambiguity is resolved by the inclusion of the initial phase offset $\phi_0$ into the spatiotemporal modulation. 
This parameter is set by the user via the sequence entered into the DMD and introduces a well-defined phase reference in the harmonic decomposition, enabling discrimination between the two conjugate solutions.
A more detailed analysis of this ambiguity and its resolution is provided in the Section S5 of the Supplementary Material.

\section{Experimental validation}

\subsection{Validation via recovering the near-field wavefront}

\begin{figure}
\centering\includegraphics[width=\textwidth]{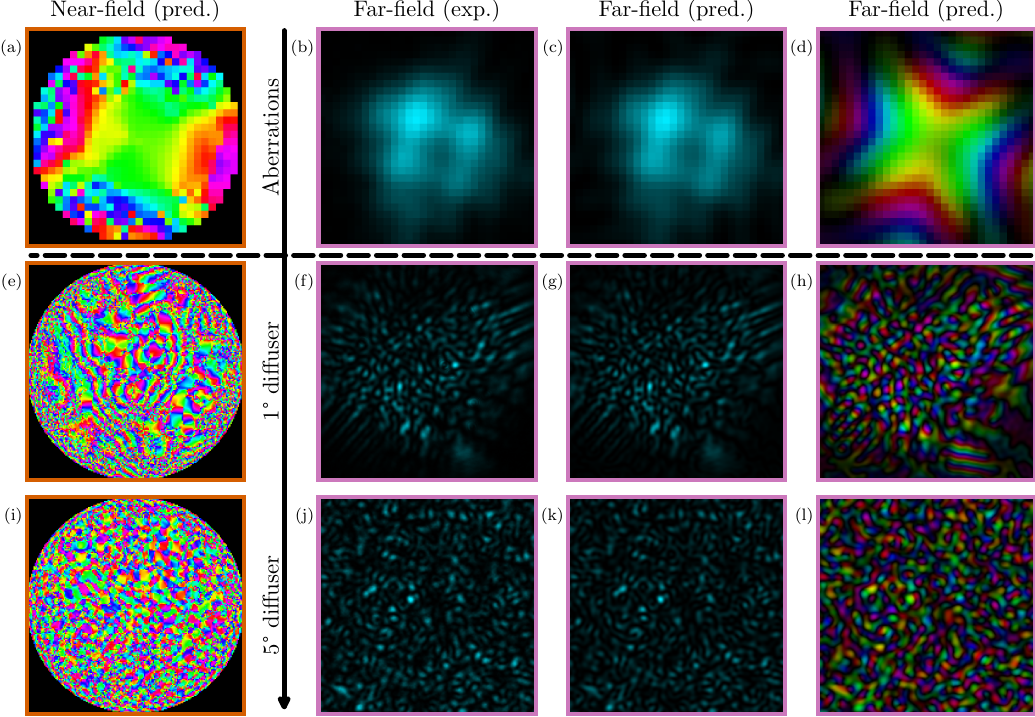}
\caption{\textbf{Experimental validation of phase retrieval across different optical regimes}.
(a), (e), and (i) Reconstructed near-field phase $D$ for (a) DMD aberrations, (e) 1$^\circ$ diffuser, and (i) 5$^\circ$ diffuser [orange: near-field plane, see Fig.\ref{fig:Modulation}(a)].
(b), (f), and (j) Experimentally measured far-field intensity distributions.
(c), (g), and (k) Predicted far-field intensity distributions obtained from the reconstructed phase.
(d), (h), and (l) Retrieved far-field phase distributions.
A strong agreement between measured and predicted intensities is observed in all cases, demonstrating accurate phase reconstruction from intensity-only measurements. All results are obtained from a single temporal acquisition (101 frames). The method remains effective across partially and fully developed speckle regimes.}
\label{fig:Results}
\end{figure}

The performance of the proposed phase retrieval algorithm is experimentally validated in three representative scenarios: 
(i) correction of low spatial frequency aberrations (intrinsic to the system and arising from the non-flat surface of the DMD~\cite{Popoff_2026}),
(ii) reconstruction of a weakly scattering diffuser (groud glass diffuser of scattering angle 1$^\circ$), 
and (iii) reconstruction of a strongly scattering diffuser (groud glass diffuser of scattering angle 5$^\circ$).
All measurements are obtained from a single temporal acquisition consisting of $1+101$ intensity frames (corresponding to the static disk illumination and the temporal sequence, respectively), from which the required observables $|a_0|^2$, $H_0$ and $H_1$ are extracted via Fourier decomposition.
The results are summarized in Fig.~\ref{fig:Results}.
Additional numerical validation of the retrieval algorithm, including topology and vortex statistics is provided in the Section S6 of the Supplementary Material.

For each experimental configuration, the retrieved near-field phase $D$, the experimentally measured far-field intensity $|a_0|^2$, and the corresponding model predictions $(a_0)_p$ are presented in Figure~\ref{fig:Results}.
The first column [Fig~\ref{fig:Results}(a), (e), and (i)] shows the reconstructed near-field phase $D$, corresponding to: (a) DMD aberrations, (e) 1$^\circ$ diffuser and (i) 5$^\circ$ diffuser.
These panels are associated with the near-field plane [orange line in Fig.~\ref{fig:Modulation}(a)], directly corresponding to the estimated phase distribution.
The second column [Fig.~\ref{fig:Results}(b), (f), and (j)] shows the experimentally measured far-field intensity distributions $|a_0|^2$, while the third column [Fig.~\ref{fig:Results}(c), (g), and (k)] shows the corresponding predicted intensities obtained from the reconstructed phase $|(a_0)_p|^2$.
These panels correspond to the far-field plane [pink line in Fig.~\ref{fig:Modulation}(a)].
An excellent agreement between measured and predicted intensity patterns is observed in all cases, indicating that the retrieved phase accurately captures the underlying optical transformation.
The fourth column [Fig.~\ref{fig:Results}(d), (h), and (l)] shows the weighted phase of the predicted far-field distributions $(a_0)_p$. 

Note that in the case of the 5$^\circ$ diffuser, the diffusion angle exceeds the numerical aperture of the optical system (See Section S7 of the Supplementaty Material). 
As a result, only a subset of the scattered spatial frequencies is collected by the detection path.
Consequently, although the algorithm accurately reconstructs the transmitted far-field phase within the accessible modal basis, 
the recovered near-field phase should be interpreted as an effective low-pass representation of the diffuser surface rather than its exact microscopic profile.
In particular, the true phase variations of the diffuser are expected to contain higher spatial frequency components than those retrieved by the reconstruction, which are not accessible due to the finite collection aperture of the system.
Nevertheless, the reconstructed phase remains sufficient to accurately reproduce the experimentally observed far-field intensity distributions.

% In the case of DMD aberrations [Fig.~\ref{fig:Results}(a)–(d)], the reconstructed near-field phase corresponds to the intrinsic wavefront distortion introduced by the device.
% The reconstruction is performed on a 30×30 grid.
% For the 1$^\circ$ diffuser [Fig.~\ref{fig:Results}(e)–(h)], the speckle pattern is partially developed. 
% Despite this, the algorithm accurately reconstructs both the near-field phase and the corresponding far-field intensity at a resolution of 240×240 pixels.
% For the 5$^\circ$ diffuser [Fig.~\ref{fig:Results}(i)–(l)], the speckle is fully developed.

The agreement between experimental measurements and model predictions is quantified using correlation coefficients summarized in Table~\ref{tab:Correlations}, 
which also reports the number of iterations to convergence, the approximate computation time, 
and the number of spatial modes for each configuration.
High correlation (> 0.98) values are observed across all configurations.
% The slight decrease in correlation for stronger scattering conditions is consistent with the increased complexity of the speckle field but does not significantly affect the reconstruction quality.
Notably, the consistently high correlation for $H_1$ demonstrates accurate recovery of phase-sensitive cross-terms, which are critical for the reconstruction.
These results confirm that the proposed harmonic-based model accurately captures the underlying optical transformation despite relying exclusively on intensity measurements.
This minimal set of observables, extracted from a single temporal acquisition, represents a significant reduction compared with conventional phase retrieval methods~\cite{Zuo_2020, Huang_2024}.

\begin{table*}
\centering
\caption{Correlation coefficients between the experimentally measured and model-predicted observables ($|a_0|^2$, $H_0$, and $H_1$) for the three experimental configurations (DMD aberrations, 1° diffuser, and 5° diffuser). The last two columns report the number of iterations required for convergence (with the approximate computation time in parentheses) and the number of spatial modes of the field.}
\label{tab:Correlations}
\begin{tabular}{|l|c|c|c|c|c|}\hline
 & $|a_0|^2$ & $H_0$ & $H_1$ & Iterations & Spatial modes \\ \hline
Aberrations & 0.9962 & 0.9977 & 0.9988 & 246 ($\sim$5\,s) & $\sim4$ \\ \hline
1° diffuser & 0.9892 & 0.9916 & 0.9904 & 489 ($\sim$38\,s) & $\sim580$ \\ \hline
5° diffuser & 0.9815 & 0.9806 & 0.9805 & 541 ($\sim$43\,s) & $\sim1900$ \\ \hline
\end{tabular}
\end{table*}

\subsection{Physical validation through wavefront correction}

\begin{figure}
\centering\includegraphics[width=\textwidth]{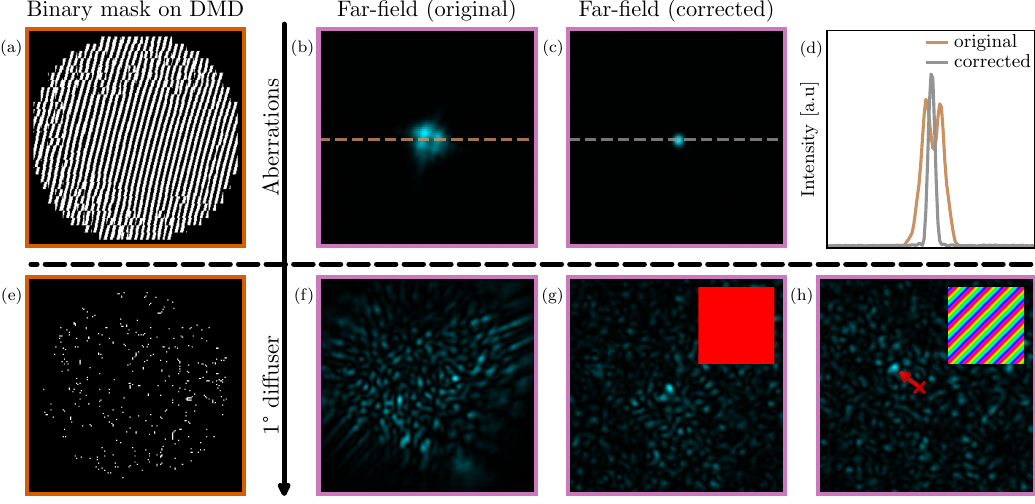}
\caption{\textbf{Physical validation of the reconstructed wavefronts through focalization}.
(a) Lee hologram encoded onto the DMD using the reconstructed near-field phase obtained from Fig.~\ref{fig:Results}(a).
(b) Experimentally measured far-field intensity before correction.
(c) Far-field intensity after applying the reconstructed phase correction through Lee holography.
(d) Transverse intensity profiles extracted along the directions indicated in (b) and (c).
(e) Representative binary mask generated from the reconstructed near-field phase obtained from the $(1^\circ)$ diffuser experiment in Fig.~\ref{fig:Results}(e).
(f) Original semi-developed speckle pattern.
(g) Far-field intensity obtained after binary phase selection.
(h) Far-field intensity obtained after introducing a linear phase ramp into the reconstructed phase prior to binary mask generation, resulting in a controlled displacement of the focal spot.
Insets in (g) and (h) show the corresponding applied phase ramps.}
\label{fig:Focus}
\end{figure}

Since the optical phase in the near field cannot be measured directly, we validate the accuracy of the retrieved phase map by using it to generate input wavefronts that compensate the phase distortions and refocus the light in the far field.
For the low spatial frequency aberrations, we generate the phase conjugated mask of the retrieved aberration phase and display it using a Lee Hologram modulation scheme onto the DMD~\cite{Lee_1979, Gutierrez-Cuevas_2024}.
However, for high spatial mode count, this approach is not possible due to the limited resolution of the DMD, we then generate a binary mask that selects regions of the diffuser with similar phase values to obtain a focalization effect in the far-field~\cite{Akbulut_2011}.
Results are summarized in Fig.~\ref{fig:Focus}.
Fig.~\ref{fig:Focus}(a) shows the binary pattern set onto the DMD corresponding to the conjugated of the retrieved aberration phase
and the experimental far-field intensities, before and after correction, are displayed in Figs.~\ref{fig:Focus}(b) and (c), respectively.
Before correction, the point spread function is strongly aberrated by the optical response of the DMD.
Contrary, the corrected far-field distribution depicts a significantly more localized focal spot.
Their transverse intensity profiles, before and after correction, are plotted in Fig.~\ref{fig:Focus}(d) confirming the improvement in focal quality after correction.

% To further validate that the retrieved phases correspond to physically meaningful optical wavefronts, additional experiments are performed using the near-field phase distributions obtained from the algorithm.
% The left column of Fig.~\ref{fig:Focus} demonstrates the correction of the intrinsic aberrations introduced by the DMD, where the reconstructed near-field phase obtained in Fig.~\ref{fig:Results}(a) is encoded by using the Lee hologram encoding technique~\cite{Lee_1979, Gutierrez-Cuevas_2024}.
% In this case, the desired phase modulation is converted into a binary amplitude hologram, where a spatial filtering selects the first diffraction order that carries the encoded phase information.
% Fig.~\ref{fig:Focus}(a) shows the corresponding binary pattern set onto the DMD, while the experimental far-field intensities, before and after correction, are displayed in Figs.~\ref{fig:Focus}(c) and (e), respectively.
% It is possible to observe that the point spread function (PSF) previous to the correction exhibits strong aberrations coming from the optical response of the DMD.
% On the other hand, the corrected far-field distribution depicts a significantly more localized focal spot.
% The transverse intensity profiles along the directions indicated in Fig.~\ref{fig:Focus}(c) and (e) are compared in Fig.~\ref{fig:Focus}(g), confirming the improvement in focal quality after correction.

For larger spatial mode counts, the same procedure cannot be directly applied due to the limited resolution of the DMD.
Instead, for the $1^\circ$ diffuser sample,a binary mask is generated that selects regions of the diffuser with similar phase values (shown in Fig.~\ref{fig:Focus}(d)), producing a focalization effect in the far field.
Although a reduced efficiency is expected compared to phase modulation~\cite{Akbulut_2011}, we observe in Fig.~\ref{fig:Focus}(g) a local enhancement of the far-field intensity compared to the original semi-developed speckle pattern shown in Fig.~\ref{fig:Focus}(f).
To confirm that the limited SNR focal spot is indeed a result of deterministic phase control rather than random constructive interference, an additional linear phase ramp is introduced into the reconstructed phase prior to generating the binary masks.
We observe in Fig.~\ref{fig:Focus}(h) that the focal spot is displaced according to the imposed phase ramp, demonstrating the validity of the retrieved near-field phase.
Note that high-spatial-frequency variations associated with the $5^\circ$ diffuser exceed the effective spatial modulation bandwidth accessible with the binary DMD encoding.
This limitation is consistent with the finite numerical aperture of the optical system discussed previously, which restricts the reconstruction to a low-spatial-frequency representation of the diffuser phase.

\section{Conclusion}

We have presented a phase retrieval framework based exclusively on intensity-only measurements and intensity-only modulations.
The method exploits the harmonic decomposition generated by spatiotemporal wavefront shaping, where a rotating angular modulation produces a set of temporal harmonics associated with spiral phase diversities.
By analyzing the harmonic content of the measured intensity signal, the proposed approach converts a phase retrieval problem into an optimization problem involving a small set of experimentally accessible observables.

The reconstruction algorithm operates directly on the measured harmonic images and does not require interferometric measurements, reference beams, phase modulators, or prior calibration of the optical system.
In particular, the reconstruction adapts to the effective numerical support of the experiment, avoiding any explicit characterization of the imaging geometry or optical aberrations.

The method was experimentally validated in three different regimes: correction of deterministic aberrations introduced by a digital micromirror device, reconstruction of a partially developed speckle field generated by a $1^\circ$ holographic diffuser, and reconstruction of a fully developed speckle field generated by a $5^\circ$ diffuser.
In all cases, the reconstructed wavefronts accurately reproduced the measured harmonic observables and generated far-field distributions in excellent agreement with the experimental measurements.

Beyond correlation coefficients agreement, the retrieved phases are shown to behave as physically meaningful optical wavefronts.
The reconstructed aberration phase enabled direct correction of the DMD point-spread function through Lee holography, while the reconstructed diffuser phase allowed deterministic focalization and controlled displacement of the focal spot through additional phase modulation.
These experiments demonstrate that the recovered phases preserve the propagation properties of the underlying optical fields.
More broadly, the proposed framework establishes a direct connection between spatiotemporal modulation, harmonic field representations, and phase retrieval.
We anticipate that this approach may provide new opportunities for wavefront sensing, calibration-free optical characterization, scattering-media imaging, and computational wavefront control in situations where only intensity measurements are available.

{\bf{Disclosures}}
The authors declare no conflicts of interest.

{\bf{Data availability}} Raw and processed data, sources to regenerate the all the figures, and sample codes for the treatment pre- and postprocessing are available in the dedicated repository openly available at the following URL/DOI: \url{https://github.com/RazoB4B/Article_PhaseRetrieval_2026.git} and in the corresponding dataset~\cite{RazoLopez_2026}.

\bibliography{bibliography}

\end{document}

\maketitle

\section{Fourier decomposition of the rotating angular mask}

\begin{figure}[h!]
\centering\includegraphics{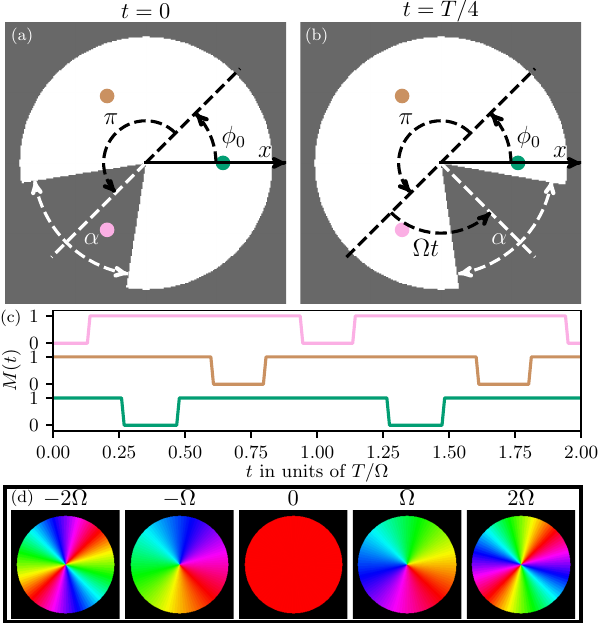}
\caption{Fourier decomposition of the rotating angular mask.
(a) Schematic representation of the rotating angular aperture $M(r, \theta, 0)$ with angular dark sector $\alpha$, rotating at angular frequency $\Omega$.
(b) Similar to (a) but considering $t=T/4$.
(c) Example temporal modulation measured at three fixed angular positions during mask rotation. Each colorful line corresponds to the temporal signal at the spatial positions indicated in panels (a) and (b).
(d) Spatial structure of the corresponding Fourier harmonics, showing the emergence of spiral-phase modes $\mathrm{e}^{\mathrm{i}n\theta}$ associated with different harmonic orders $n$.}
\label{fig:Masks}
\end{figure}

To understand the origin of the spiral-phase harmonics observed experimentally, we analyze the temporal Fourier decomposition of the rotating angular mask introduced in the main text.
The modulation function is defined as
\begin{equation}
    M(r, \theta, t) = \left\{\begin{array}{l l} 0, & r\geq r_{\max} \\
    0, & r<r_{\max}, \quad \mathrm{and} \quad \left|\theta - \theta_c(t)\right|<\alpha/2 \\
    1, & \mathrm{otherwhise}
    \end{array} \right.
\end{equation}
where $\theta_c(t)=\Omega t + \phi_0 + \pi$, $\alpha$ is the angle of the dark rotating sector, $\Omega$ is the angular rotation frequency, and $\phi_0$ is the initial phase offset.
Schematic representations of the rotating mask are shown in Figs.~\ref{fig:Masks}(a) and (b) for $\alpha=2\pi/5$.
Additionally, Fig.~\ref{fig:Masks}(c) plots the amplitude variation with respect to time for three spatial points in the time-modulating mask. 

Because the modulation is periodic in time, it can be expanded into a temporal Fourier series
\begin{equation}
    M(r, \theta, t) = \sum_{n=-\infty}^{\infty} c_n(\theta)\mathrm{e}^{\mathrm{i}n\Omega t}
\end{equation}
with Fourier coefficients
\begin{equation}
    c_n(\theta) = \frac{1}{T} \int_{0}^{T} M(r, \theta, t) \mathrm{e}^{-\mathrm{i}n\Omega t},
\end{equation}
where $T$ is the period of one lap.
By evaluating the integral over the angular support of the mask, the coefficients become
\begin{equation}
    c_n(\theta) \propto \mathrm{sinc}\left(\frac{n \alpha}{2}\right)\mathrm{e}^{\mathrm{i}n (\theta-\phi_0 - \alpha/2)} \mathrm{d}t,
\end{equation}
leading to the harmonic decomposition
\begin{equation}
    P(\theta, \Omega) \propto \delta(\Omega) + \sum_{n \neq 0}\mathrm{sinc}\left(\frac{n \alpha}{2}\right)\mathrm{e}^{\mathrm{i}n (\theta-\phi_0 - \alpha/2)}\delta(n\Omega), \label{eq:frqSpectrum}
\end{equation}
The decomposition reveals that each temporal harmonic is associated with a spatial spiral-phase term of the form $\mathrm{e}^{\mathrm{i}n\theta}$ corresponding to a vortex mode with topological charge $n$.
The harmonic amplitudes are weighted by the sinc envelope $\mathrm{sinc}(n\alpha/2)$ which selectively suppresses specific harmonic orders depending on the angular aperture $\alpha$. 
For instance, the cases of $\alpha=\pi$, where are even harmonics are suppressed, or $\alpha=2\pi/3$ that suppresses harmonics multiple of three are experimentally verified in Fig. 2 of the main text.
Figure~\ref{fig:Masks}(d) illustrates the relation between the rotating angular modulation, the temporal harmonic decomposition, and the resulting spiral-phase modes.

\section{Experimental setup}

\begin{figure}[h!]
\centering\includegraphics{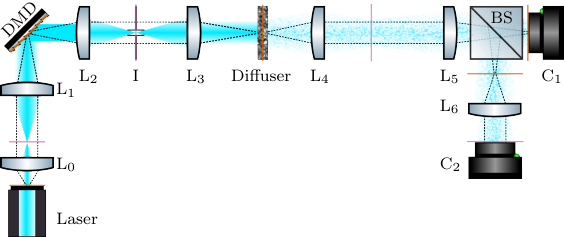}
\caption{Setup for spatiotemporal wavefront shaping.
Orange markers indicate near-field planes, while pink markers indicate far-field planes, following the convention used throughout the manuscript.
An iris placed inside the relay telescope is used for spatial filtering during Lee hologram generation.
Distances are not to scale.
C: CCD camera, L: lenses, BS: Beam splitter, DMD: digital micromirror device, I: adjustable iris.}
\label{fig:Setup}
\end{figure}

The complete optical setup used throughout the experiments is shown in Fig.~\ref{fig:Setup}.
A coherent laser beam (Coherent Sapphire SF NX \@ 488nm) is first expanded using a 4f telescope before illuminating the Vialux digital micromirror device (DMD), which generates the time-dependent angular modulation patterns described in the main text.
The modulated field is relayed through a second 4f system toward the diffuser plane.
An iris is positioned within the relay telescope between the DMD and the diffuser plane in order to control the spatial frequency content of the illumination.
This element is particularly relevant for the Lee hologram experiments discussed in the main text, where spatial filtering is used to isolate the desired diffraction order generated by the DMD.

Depending on the experiment, the diffuser corresponds either to the intrinsic aberration profile of the DMD or to an external holographic diffuser (Edmund with scattering angle 5$^\circ$ or Thorlabs with scattering angle 1$^\circ$).
After propagation through the diffuser, the optical field is imaged onto the detection stage using an additional 4f system. 
A beam splitter located before the final imaging lens separates the beam into two detection paths: a near-field imaging path, and a far-field imaging path implemented through a 2f configuration.

The corresponding detection planes are identified throughout the manuscript using the same color convention introduced in Fig. 2 of the main text: orange lines indicate near-field planes, pink lines indicate far-field (conjugated) planes.
The cameras used for acquisition are labeled C$_1$ and C$_2$, while the lenses are labeled L$_i$, with $i=0,\dots,6$.
This configuration enables simultaneous access to near-field and far-field intensity distributions while preserving a common spatiotemporal modulation scheme for all experiments presented in the manuscript.

\section{Global scaling parameter}

The forward model used by the reconstruction propagates the current diffuser estimate $D$ to the detection plane through a discrete Fourier transform.
The mapping between the near-field and far-field sampling grids depends on physical parameters of the imaging system, such as the effective focal length, the magnification, the pixel size, and the numerical aperture, which are generally not known with sufficient accuracy.
Rather than calibrating these quantities explicitly, we introduce a single global scaling parameter $\gamma$ that rescales the spatial coordinates of the predicted far fields, resampling the model output so that its effective spatial support matches that of the measured observables.
This parameter is treated as an additional unknown and optimized jointly with the diffuser phase $D$ by gradient descent, using the same normalized $L^2$ loss as in the main text.
Because it absorbs the unknown sampling of the system, $\gamma$ makes the method applicable to partially characterized optical configurations and to partially developed speckle regimes, where the effective numerical support is not known a priori.

\section{Multiscale strategy}

When considering large images ($>60\times60$ pixels), the model uses a multiscale strategy based on $m$ different stages.
At different stages, the algorithm only takes into account center-cropped versions of the three different experimental observables.
Such cropped far-field images have an initial size of $d/2^m$ during the first stage, where $d$ is the dimension of the original images and $m$ is an integer that is successively decreased by one until reaching zero and that is fixed by the user.
Similarly, the reconstructed near-field phase is initialized on a grid of size $d^\prime/2^m$, where $d^\prime$ denotes the user-defined reconstruction size of the diffuser. 
At each scale transition, the reconstructed near-field phase is upsampled to initialize the next optimization stage, while the corresponding far-field observables are replaced by larger center-cropped images.
Consequently, every new stage introduces progressively higher spatial frequencies into the reconstruction while preserving the low-frequency solution obtained at the previous level.
A patience parameter (typically 100 iterations) determines when convergence has been reached before advancing to the next resolution level.
Upon convergence ($m=0$), the algorithm returns the reconstructed near-field phase $D$ and the corresponding complex far-field $a_0$.

\section{Symmetry ambiguity and solvability conditions}

The proposed reconstruction framework is subject to the intrinsic conjugation ambiguity commonly encountered in phase retrieval problems. 
In particular, the diffuser distributions $D(r)$ and $D^*(-r)$ produce complex conjugated patterns, {\textit{i.e.}} identical intensity measurements, and therefore cannot be distinguished using intensity-only observables alone.
In the present approach, this ambiguity is resolved through the inclusion of the initial phase $\phi_0$, which introduces a phase offset in the harmonic coefficients $\mathrm{sinc}\left(n \alpha/2\right)\mathrm{e}^{\mathrm{i}n\phi_0}$, thereby breaking the conjugation symmetry between the two solutions.
Indeed, under the transformation $D(r)\rightarrow D^*(-r)$ the harmonic components acquire opposite phase factors through the sign inversion of the Fourier coefficients, allowing the reconstruction algorithm to discriminate between both solutions.

\begin{figure}[h!]
\centering\includegraphics{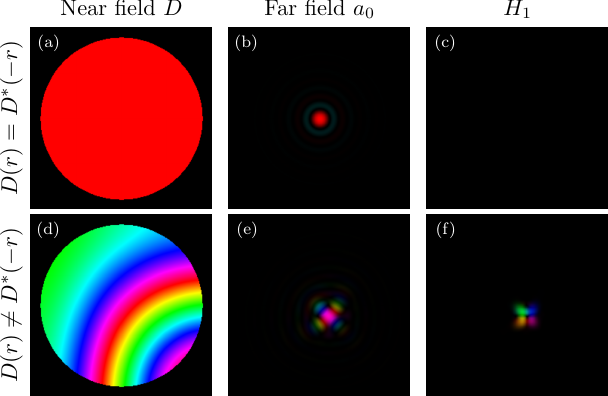}
\caption{Symmetry ambiguity and solvability conditions of the reconstruction problem.
(a)–(c) Symmetric optical system satisfying $D(r)=D^*(-r)$.
The resulting far-field phase is spatially uniform and the first harmonic vanishes $H_1=0$, preventing unique phase reconstruction.
(d)–(f) Slightly asymmetric system obtained by introducing weak aberrations.
The symmetry breaking generates nonzero harmonic information $H_1\neq0$, enabling successful reconstruction of the optical phase.}
\label{fig:Ambiguity}
\end{figure}

A particular class of systems remains fundamentally ambiguous.
Specifically, when the diffuser satisfies the symmetry condition $D(r)=D^*(-r)$, the resulting far-field phase becomes spatially uniform and the harmonic components satisfy $a_1=-a^*_{-1}$ leading to the cancellation of the first harmonic observable $H_1=0$.
In this situation, the measurements themselves do not contain sufficient information to uniquely recover the phase distribution. 
Therefore, the absence of reconstruction is not a limitation of the optimization algorithm, but rather a consequence of the symmetry of the optical system.
This behavior is illustrated by Fig.~\ref{fig:Ambiguity}.
The upper row corresponds to a symmetric optical system producing a flat far-field phase and a vanishing first harmonic $H_1$.
In contrast, the lower row shows that even weak aberrations break the conjugation symmetry, generating nonzero harmonic information and enabling successful reconstruction.
This behavior is consistent with the experimental observations presented in Fig. 4 of the main manuscript, where the intrinsic aberrations of the DMD introduce sufficient asymmetry to render the system solvable.

\section{Numerical validation of the retrieval algorithm}

The reconstruction accuracy of the proposed method is quantitatively evaluated by performing numerical simulations using synthetic diffuser phase distributions with known ground truth. 
The retrieved near-field and far-field quantities are then compared against the original near- and far-fields.

\subsection{Field reconstruction accuracy}

\begin{figure}[h!]
\centering\includegraphics{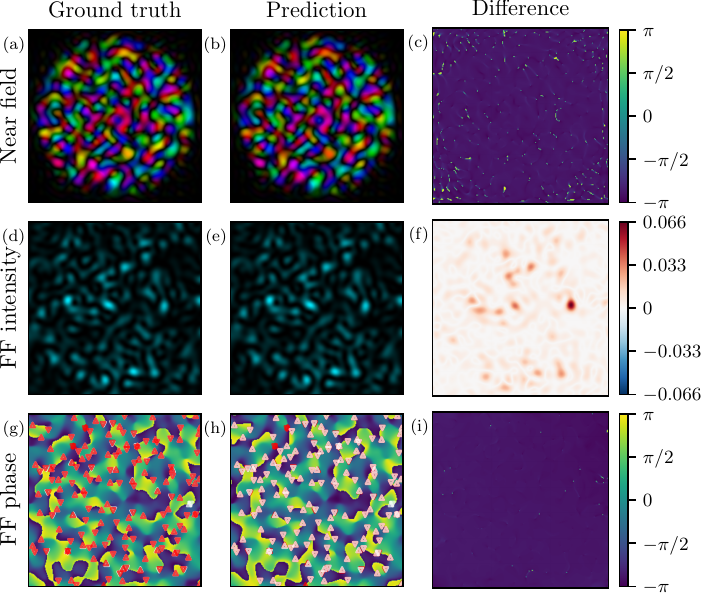}
\caption{Numerical reconstruction of near- and far-field wavefronts.
(a) Ground-truth near-field wavefront.
(b) Reconstructed near-field wavefront.
(c) Phase difference between the original and reconstructed near-field phases.
(d) Ground-truth far-field intensity distribution.
(e) Reconstructed far-field intensity distribution.
(f) Absolute residual map between the original and reconstructed far-field intensities.
(g) Ground-truth far-field phase distribution.
(h) Reconstructed far-field phase distribution.
(i) Phase difference between the original and reconstructed far-field phases.
Positive and negative phase singularities are represented in (g) and (h) by upward and downward triangles, respectively. 
Original vortices are shown in white, while reconstructed vortices are superimposed in red.}
\label{fig:NumRecons}
\end{figure}

The original and reconstructed optical fields obtained from the numerical simulations are compared in Fig.~\ref{fig:NumRecons}.
Specifically, Fig.~\ref{fig:NumRecons}(a) and (b) plot the ground truth and predicted near-fields, $D$ and $D_p$ respectively, while Fig.~\ref{fig:NumRecons}(c) presents their phase difference.
The reconstruction accurately reproduces the original diffuser phase up to an approximately uniform phase offset with a correlation coefficient of 0.9986.
The corresponding far-field intensity distributions are shown in Fig.~\ref{fig:NumRecons}(d) and (e), together with the absolute residual map in Fig.~\ref{fig:NumRecons}(f).
The residual map is computed as $\left||a_0|^2-|(a_0)_p|^2\right|$, where $a_0$ and $(a_0)_p$ are the normalized ground truth and reconstructed far-field wavefront, respectively.
Finally, Figs.~\ref{fig:NumRecons}(g) and (h) show the original and reconstructed far-field phase distributions $\phi$ and $\phi_p$, and Fig.~\ref{fig:NumRecons}(i) plots their difference $\phi-\phi_p$.
The nearly flat profile confirms that the reconstruction error is dominated by a global phase factor.
The reconstructed far-field wavefront $(a_0)_p$ yields to a correlation coefficient of 0.9985 with respect to the ground truth $a_0$.

\subsection{Cross-correlation analysis}

\begin{figure}[h!]
\centering\includegraphics{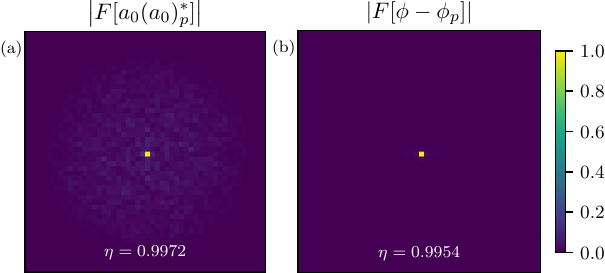}
\caption{Cross-correlation analysis of the reconstructed fields.
(a) Absolute value of the cross-correlation spectrum computed from the complex far-field amplitudes.
(b) Cross-correlation spectrum obtained after removing intensity fluctuations from the phase distributions.}
\label{fig:CCSpectra}
\end{figure}

To further quantify the reconstruction quality, the cross-correlation spectra between the original and reconstructed fields are evaluated.
Figure~\ref{fig:CCSpectra}(a) shows the absolute value of the cross-correlation spectrum computed using the complex far-field amplitudes $|F[a_0(a_0)_p^*]|$, where $F[\cdot]$ stands for the Fourier transform.
The fraction of optical energy preserved in the correlated coherent speckle mode $\eta$ can be estimated by the value of the central peak ($>0.99$ in this case), indicating a strong spatial correlation between the reconstructed and target fields.
Here, the pale broad halo indicates an incoherent or partially correlated intensity structure.

Since the central peak is influenced by both, phase and amplitude, correlations, a high full-field peak does not necessarily mean the phases match perfectly.
Then, the wavefront similarity can be isolated from intensity fluctuations by computing a phase-only cross-correlation spectrum.
Figure~\ref{fig:CCSpectra}(b) plots the corresponding spectrum after removing intensity fluctuations from the phase distributions $|F[\mathrm{e}^{\left( \phi-\phi_p \right)}]|$. 
In this case, the diffuse speckle background around the central peak is strongly suppressed.
Identically to Fig.~\ref{fig:CCSpectra}(a), the central peak allows to estimate that more of the 99.9\% of the total energy is preserved, demonstrating an almost perfect reconstruction of the phase structure.

\subsection{Topological validation through phase singularities}

As a last step, the topology of the retrieved phase field is evaluated by identifying the singularities (optical vortices) in both the original and reconstructed far-field phase distributions. 
In Fig.~\ref{fig:NumRecons}(g) and (h), positive and negative vortices are represented by upward and downward triangles, respectively.
Specifically, the vortices of the reconstructed field are superimposed in red over the original singularity distribution in Fig.~\ref{fig:NumRecons}(g), while Fig.~\ref{fig:NumRecons}(h) shows the inverse comparison.
The reconstructed field preserves nearly all singularity positions and charges, with only two additional vortex-antivortex pairs appearing during the reconstruction process.

\begin{figure}[h!]
\centering\includegraphics{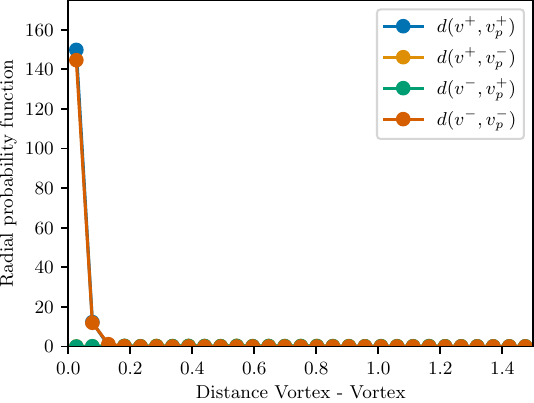}
\caption{Statistical analysis of phase singularity reconstruction.
Radial probability distributions for vortex matching between the original and reconstructed far-field phases as a function of normalized distance.
Equal-charge correlations exhibit strong localization at zero distance, while opposite-charge correlations remain negligible.}
\label{fig:Vortex}
\end{figure}

In figure~\ref{fig:Vortex}, we plot the radial probability distributions for vortex matching as a function of the distance normalized by the average speckle size between the original and reconstructed far-field phases.
Distributions are computed by considering 10 different wavefront reconstructions of speckle size $\sim12$ pixels.
The distributions $d(v^+, v^+_p)$ and $d(v^-, v^-_p)$ corresponds to equal-charge vortex matching, and exhibit sharp peaks at zero distance, demonstrating that nearly all vortices are accurately recovered.
In contrast, the cross-charge distributions $d(v^+, v^-_p)$ and $d(v^-, v^+_p)$ remain negligible for all distances, confirming the conservation of vortex charge during reconstruction.
From these distributions and considering the speckle size, we estimate that approximately 99\% of the phase singularities are correctly recovered within a one-pixel tolerance.

\section{Effect of the finite numerical aperture}

\begin{figure}[h!]
\centering\includegraphics{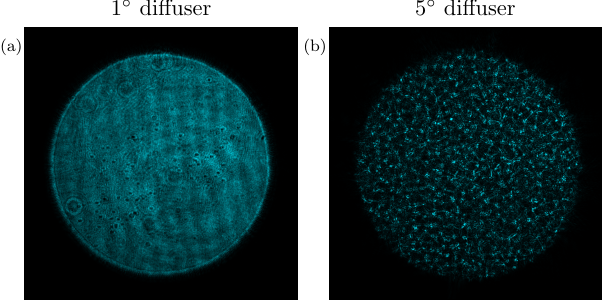}
\caption{Experimental near-field intensity distributions for different diffuser scattering angles.
(a) Near-field intensity measured for the $1^\circ$ diffuser.
(b) Near-field intensity measured for the $5^\circ$ diffuser.}
\label{fig:NearFields}
\end{figure}

As discussed in the main text, the reconstruction obtained for the $5^\circ$ diffuser must be interpreted as an effective low-pass representation of the transmitted wavefront due to the finite numerical aperture of the optical system. 
To illustrate this effect experimentally, Fig.~\ref{fig:NearFields} compares the near-field intensity distributions measured immediately after the diffuser for the (a) $1^\circ$ and  (b) $5^\circ$ diffusers.
For the $1^\circ$ diffuser, the near-field intensity remains approximately uniform over the illuminated aperture, indicating that all the scattered spatial frequencies are collected by the imaging system.
Under these conditions, the measured field provides a faithful representation of the optical wavefront transmitted through the diffuser.
In contrast, the near-field intensity obtained with the $5^\circ$ diffuser exhibits pronounced intensity fluctuations and regions of reduced or null illumination. 
This behavior indicates that a significant fraction of the scattered light falls outside the collection numerical aperture of the optical system.
Consequently, only a subset of the transmitted spatial frequencies contributes to the measured field.
As a result, the phase retrieved by the reconstruction algorithm corresponds to the effective wavefront supported by the collected modes rather than to the complete microscopic phase profile of the diffuser.
This observation is consistent with the discussion presented in the main text, where the recovered near-field phase for the $5^\circ$ diffuser is interpreted as a low-spatial-frequency representation of the underlying diffuser surface.
Importantly, although high spatial frequencies are lost due to numerical-aperture truncation, the recovered phase remains sufficient to accurately reproduce the experimentally observed far-field intensity distributions.

